\documentclass[aps,twocolumn,superscriptaddress]{revtex4}
\usepackage[usenames,dvipsnames]{color}
\usepackage{graphicx}
\usepackage{amsmath}
\usepackage{amssymb}

\newcommand{\symkowm}[2]{ \left<\left[ #1 ; #2 \right]_{+} \right>}
\begin{document}
\title{Quantum partition of energy for a free Brownian particle: \\Impact of dissipation}
%
\author{J. Spiechowicz}
\affiliation{Institute of Physics and Silesian Center for Education and Interdisciplinary Research, University of Silesia, 41-500 Chorz{\'o}w, Poland}
\author{P. Bialas}
\affiliation{Institute of Physics and Silesian Center for Education and Interdisciplinary Research, University of Silesia, 41-500 Chorz{\'o}w, Poland}
\author{J. {\L}uczka}
\affiliation{Institute of Physics and Silesian Center for Education and Interdisciplinary Research, University of Silesia, 41-500 Chorz{\'o}w, Poland}
\email{jerzy.luczka@us.edu.pl}

\begin{abstract}
We study the quantum counterpart of the theorem on energy equipartition for classical systems.  We consider  a free quantum Brownian particle modelled in terms of the Caldeira-Leggett framework: a system plus thermostat consisting of an infinite number of harmonic oscillators. By virtue of the theorem on the averaged kinetic energy $E_k$ of the quantum particle, it  is expressed as $E_k = \langle \mathcal E_k \rangle$, where $\mathcal E_k$ is thermal kinetic energy of the thermostat per one degree of freedom and $\langle ...\rangle$ denotes averaging over frequencies $\omega$ of thermostat oscillators which contribute to $E_k$ according to the probability distribution $\mathbb P(\omega)$.  We explore the impact of various dissipation mechanisms,  via the  Drude, Gaussian, algebraic and Debye spectral density functions, on the characteristic features of $\mathbb{P}(\omega)$.     The role  of the system-thermostat coupling strength and the memory time on the most probable thermostat oscillator frequency as well as the kinetic energy $E_k$ of the Brownian particle is analysed.  
\end{abstract}

\maketitle

\section{Introduction}
Quantum physics shows that its world can exhibit behavior which is radically different than its classical counterpart. Wave-particle duality, entanglement of states,  decoherence, Casimir force, quantum information -  there are generic  examples which in turn carry the potential for new applications in the near or 
further future. Yet, there still remain new properties, behavior and  phenomena to be uncovered for this world. In this context, the quantum counterpart of the theorem on energy equipartition  for classical systems is still not  formulated in a general case. We attempt to make one step forward. 
In  classical statistical physics,  the  theorem  on equipartition of energy states that for a system in thermodynamic equilibrium its kinetic
energy $E_k$  is shared equally amongst all energetically-accessible degrees of  freedom. It also  relates average energy $E_k=k_BT/2$ per one degree of freedom to the temperature $T$ of the system ($k_B$ is the Boltzmann constant). When thermostat is modelled as an infinite collection of harmonic oscillators of temperature $T$ then the averaged kinetic energy of thermostat per one degree of freedom is also $\mathcal E_k=k_BT/2$. In other words, $E_k=\mathcal E_k$ and all degrees of freedom of both system and thermostat have exactly the same averaged kinetic energy. This is why it is named 'equipartition'.  It is universal in the sense that it does not depend on a number of particles in the system, a potential force which acts on them, any interaction between particles or the strength of coupling between the system and thermostat \cite{huang,ter}. 
For quantum systems, in a general case its counterpart is not known. In literature, one can find reports on energetics  of selected quantum systems \cite{hakim}. In Ref. \cite{ford85}, an exact expression for the free energy of a quantum oscillator interacting, via dipole coupling, with a blackbody radiation field was derived. Next, the same authors studied the similar problem by more conventional method using the fluctuation-dissipation theorem  and obtained the expression for kinetic energy of the quantum oscillator \cite{ford88}. At the same time, the review on quantum Brownian motion  was published \cite{ingol1}. Formulas for the variance  of position and momentum of the oscillator is presented in the Table 2 therein. 
There are also books \cite{weis,landau,zubarev,breuer} in which different versions of  kinetic energy of a free Brownian particle  can be obtained directly or indirectly. 
Lately, kinetic energy of a trappped Fermi gas has been considered \cite{grela}.  Many other aspects of quantum Brownian motion has been intensively studied in last few years \cite{lewenstein,korbicz,smirne,ankerhold,editor,china, carlesso, lampo,lampo2}. However, the previous results have not been {\it directly}  related to  the energy  equipartition theorem.  Very recently,  some progress  has been made  in formulation of this law  assuming that thermostat is a collection of an infinite number of quantum oscillators  \cite{bialas,arxiv2018}.  In contrary to the classical case, the averaged kinetic energy of the thermostat oscillator depends on its frequency, $\mathcal E_k = \mathcal E_k(\omega)$, and in a consequence the kinetic energy of the Brownian particle $E_k$ depends on all $\mathcal E_k$ but in a non-uniform way determined by a probability distribution $\mathbb{P}(\omega)$ of the thermostat oscillator frequencies $\omega$. In turn, $\mathbb{P}(\omega)$ depends on  microscopic details of thermostat and interactions.  The latter aspect can be modeled  by the spectral density of thermostat modes which contains necessary   information on the system-thermostat interaction. The aim of this work is to analyse the impact of various dissipation mechanisms on the kinetic energy $E_k$ of the free Brownian particle.

The paper is structured as follows. The presentation starts in Section II, where for the paper to be self-contained we recapitulate very briefly some of the well known key points on the quantum Brownian motion. We apply a simple yet powerful minimal model based on the concept of Hamiltonian for a composite quantum system: a Brownian particle-plus-thermostat \cite{maga}. Starting from the Heisenberg equations of motion for all position and momentum operators, an exact effective evolution equation can be derived for the coordinate and momentum operators of the Brownian particle. This integro-differential equation is called a generalized quantum Langevin equation in which an integral (damping) kernel and a thermal noise term are related via the fluctuation-dissipation theorem.  We recall a solution of this equation for the momentum  of the free Brownian particle and present  the quantum law for energy  partition of  the Brownian particle which has been derived in Ref. \cite{arxiv2018}.
In Section III, we comment on the energy partition theorem and discuss on relations to the fluctuation-dissipation theorem derived in the linear response theory. 
In the main part of the paper, in Section  IV,  we are interested in the impact of various dissipation mechanisms on  $\mathbb P(\omega)$.  This mechanism is modelled via the damping kernel of the Langevin equation. We consider two families of the memory functions: (i) exponentially and (ii) algebraically decaying. Two sub-families are analyzed: (a) monotonically and (b) periodically decaying functions. It covers majority of crucial and accessible models of dissipation mechanisms.
On one hand, we reveal similarities for the impact of various dissipation mechanisms and, on the other one, there are interesting and significant differences. 
In Section V, we analyse the first two statistical moments of the frequency probability distribution. The first moment  is directly related to the averaged kinetic energy at zero temperature while the second moment -  the first quantum correction to the classical result in the high temperature regime. 
We summarise the results of the work in the last Section VI. In Appendices we present the solution of the generalized Langevin equation, derive the formula for the kinetic energy of the Brownian particle and present the fluctuation-dissipation relation.

\section{Partition of energy for a free Brownian particle}
An archetype of Brownian motion of a quantum particle is based on the Hamiltonian description of a composite system: the quantum particle-plus-thermostat. By way of explanation, the particle  of mass $M$ is subjected to the potential $U(x)$ and interacts with a large number of independent oscillators, which form a thermal reservoir of temperature $T$.  
 The typical quantum-mechanical Hamiltonian of such a closed (and conservative) system assumes the form    $\grave a\  la$ Caldeira-Leggett ones \cite{maga,uler,caldeira,ford,gaussian,van,et2,ph,weis,chaos}: 
\begin{equation}
H=\frac{p^2}{2M}+U(x) + \sum_i \left[ \frac{p_i^2}{2m_i} + \frac{m_i
\omega_i^2}{2} \left( q_i - \frac{c_i}{m_i \omega_i^2} x\right)^2 \right].  
\end{equation}
The coordinate and momentum operators $\{x, p\}$ refer to the Brownian particle and $\{q_i, p_i\}$ are the coordinate and momentum operators of the $i$-th heat bath oscillator of mass $m_i$ and the eigenfrequency $\omega_i$. The parameter $c_i$ characterizes the interaction strength of the particle with the $i$-th oscillator. There is the counter-term, the last term proportional to $x^2$, which is included to cancel a harmonic contribution to the particle potential. All coordinate and momentum operators obey canonical equal-time commutation relations. 
 
The next step is to write the Heisenberg equations of motion for all coordinate and momentum operators $\{x, p, q_i,p_i\}$ and solve Heisenberg equations for the reservoir operators to obtain an effective equation of motion only for the particle coordinate $x(t)$. It is the so-called generalized quantum Langevin equation which reads (for detailed derivation, see e.g. \cite{bialas})
\begin{equation}\label{GLE}
M{\ddot x}(t)
+\int_0^t \gamma(t-s) \dot{x}(s) \, ds = -U'(x(t)) -\gamma(t) x(0)+ \eta(t), 
\end{equation}
where $\dot{x}(t)=p(t)/M$, $U'(x)$ denotes differentiation with respect to $x$, 
 $\gamma(t)$ is a dissipation function (damping or memory kernel), 
\begin{equation} \label{diss}
\gamma(t) =\sum_i \frac{c_i^2}{m_i \omega_i^2} \cos(\omega_i t)  \equiv 
\int_0^{\infty} d \omega J(\omega) \cos(\omega t), 
\end{equation}
where 
\begin{eqnarray} \label{spectral}
J(\omega) = \sum_i \frac{ c_i^2}{ m_i \omega_i^2} \delta(\omega -\omega_i) 
\end{eqnarray} 
is a spectral function of a heat bath which contains information on its modes and the system-heat bath interaction. The term $\eta(t)$ can be interpreted as a random force acting on the Brownian particle, 
\begin{equation} 
\eta(t) =\sum_i c_i \left[q_i(0) \cos(\omega_i t) + \frac{p_i(0)}{m_i \omega_i}\sin(\omega_i t) \right]
\label{force} 
\end{equation}
It  depends  on the initial conditions imposed on oscillators of the thermostat.  We note that  effective dynamics of the quantum Brownian particle is described by an integro-differential equation for the coordinate operator $x(t)$ and the initial condition $x(0)$ occurs in this evolution equation. It is something untypical for ordinary differential equations. Usually, the initial conditions are separated from equations of motion and independently accompanied to them. Here, for the open system, the initial conditions are an integral part of the effective  dynamics and not an independent input.   The initial preparation of the total system fixes statistical properties of the thermostat and the Brownian particle. 

We consider the free Brownian particle for which $U'(x)=0$. From Eq. (\ref{GLE}) one obtains the equation of motion for the momentum operator, 
\begin{equation}\label{GLE2}
{\dot p}(t)
+\frac{1}{M} \int_0^t \gamma(t-s) p(s) \, ds = -\gamma(t) x(0)+ \eta(t). 
\end{equation}
Its solution reads (see Appendix A)
\begin{align}\label{p(t)} 
p(t)& = R(t)p(0) - \int_0^t du\; R(t-u) \gamma(u)x(0)\nonumber \\  &+ \int_0^t du\; R(t-u) \eta(u),  
\end{align}
where $R(t)$ is a response function determined by its Laplace transform,  
\begin{equation}\label{RL} 
\hat{R}_L(z) = \frac{M}{Mz + \hat \gamma_L(z)}. 
\end{equation}  
Here, $\hat \gamma_L(z)$ is a Laplace transform of the dissipation function $\gamma(t)$ and for any function  $f(t)$ its Laplace transform is defined as 
\begin{equation}\label{fL} 
\hat f_L(z) = \int_0^{\infty} dt \; {\mbox e}^{-zt} f(t). 
\end{equation}  
Using Eq. (\ref{p(t)}), one can  calculate averaged kinetic energy 
$E_k(t)=\langle p^2(t)\rangle/2M$ of the Brownian particle. In the long time 
limit $t \to \infty$, when a thermal equilibrium state is reached, it has the form (see Eq. (\ref{Ek2}) in Appendix B)
\begin{equation}\label{Ek}
E_k = \langle \mathcal{E}_k \rangle = \int_0^{\infty} d\omega \; \mathcal{E}_k(\omega)\mathbb{P}(\omega),  
\end{equation}
where  
\begin{equation}\label{ho}
\mathcal{E}_k(\omega) = \frac{\hbar \omega}{4} \coth\left({\frac{\hbar \omega}{ 2k_BT}}\right)
\end{equation} 
is thermal kinetic energy per one degree of freedom of the thermostat consisting of free harmonic oscillators \cite{feynman} and $\langle ...\rangle$ denotes averaging over frequencies $\omega$ of those thermostat oscillators which contribute to $E_k$ according to the probability distribution (see Eq. (\ref{P2}) in Appendix B)
\begin{eqnarray}\label{P}
\mathbb{P}(\omega) = \frac{1}{\pi} \left[\hat{R}_L(i\omega) + \hat{R}_L(-i\omega) \right]. 
\end{eqnarray}
The formula (\ref{Ek}) together with Eq. (\ref{P}) constitutes a \emph{quantum law for partition of energy}. It means that the averaged  kinetic energy $E_k$ of the Brownian particle is an averaged kinetic energy $\mathcal E_k$ per one degree of freedom of the thermostat oscillators. The averaging is twofold:  (i) over the thermal equilibrium Gibbs state for the thermostat oscillators resulting in $\mathcal{E}_k(\omega)$ given by \mbox{Eq. (\ref{ho})} and (ii) over frequencies $\omega$ of  those thermostat oscillators which contribute to $E_k$  according to the probability distribution   $\mathbb P(\omega)\ge 0$ which is normalized on the frequency half-line \cite{arxiv2018}:  $\int_0^{\infty} d\omega \; {\mathbb P}(\omega) = 1.$

We rewrite the formula (\ref{P}) to the form which is convenient for calculations. To this aim we note that the Laplace transform can be expressed by the cosine  and  sine Fourier transforms. In particular, 
\begin{subequations}
\begin{align} 
\hat{\gamma}_L(i\omega) &= \int_0^{\infty} dt \, \gamma(t)  \mbox{e}^{-i\omega t}  = A(\omega) - i B(\omega) \label{L-F}\\
A(\omega) &= \int_0^{\infty} dt \; \gamma(t) \cos{(\omega t)}, \label{cos}\\  
B(\omega) &= \int_0^{\infty} dt \; \gamma(t) \sin{(\omega t)}. \label{sin}
\end{align}
\end{subequations}
We put it into Eqs. (\ref{RL}) and (\ref{P}) to get the following expression:
\begin{equation} \label{Pp}
\mathbb{P}(\omega) = \frac{2 M}{\pi} \frac{ A(\omega)}{A^2(\omega)+[B(\omega)-M\omega]^2}.
\end{equation}
Let us observe that the function $A(\omega)$ is related to the spectral function $J(\omega)$. Indeed, from Eqs. (\ref{diss}), (\ref{inverseG}) in Appendix C and  the definition (\ref{cos}) of $A(\omega)$ it follows that   $A(\omega)=(\pi/2) J(\omega)$.  Because the spectral function (\ref{spectral}) is non-negative, $J(\omega) \ge 0$, and the denominator in (\ref{Pp}) is positive, the function $ \mathbb{P}(\omega)$ is non-negative as required. 

The representation (\ref{Pp}) allows to study the influence of various forms of the dissipation function $\gamma(t)$ or equivalently the spectral density $J(\omega)$.
\begin{figure*}[t]
	\centering
    \includegraphics[width=0.39\linewidth]{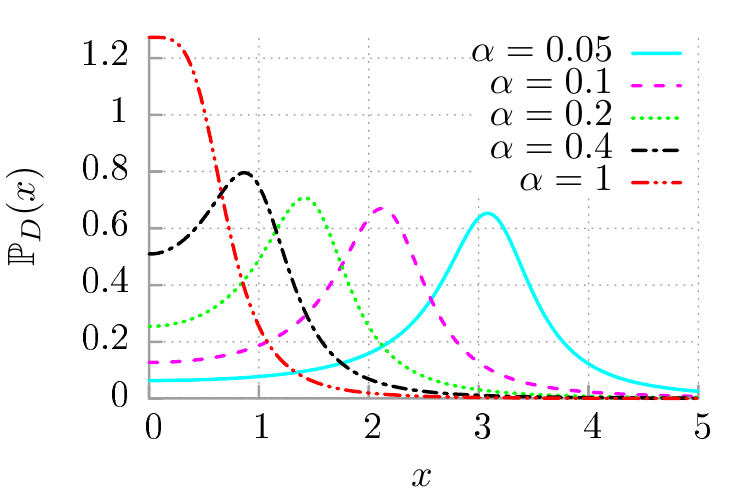}
    \includegraphics[width=0.39\linewidth]{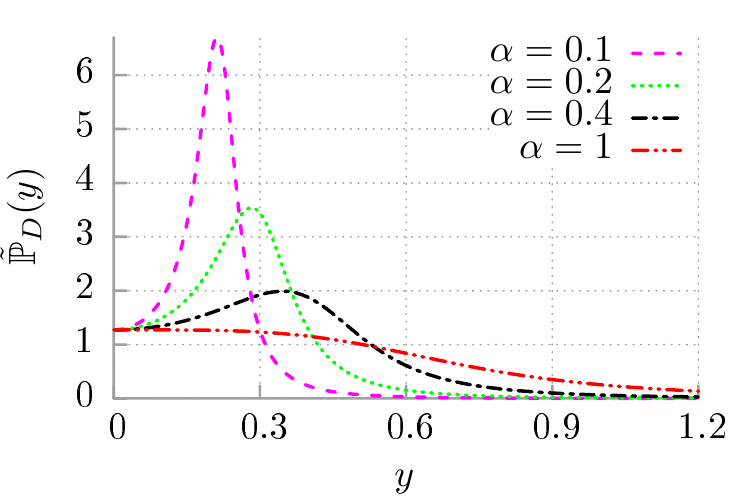}
    \caption{(color online): Exponential decay of the dissipation function \mbox{$\gamma_D(t) = (\gamma_0/2 \tau_c)e^{-t/\tau_c}$} known as the Drude model. Probability distributions $\mathbb{P}_D(x)$ and $\tilde{\mathbb{P}}_D(y)$ in two different scalings are shown for selected values of the dimensionless parameter $\alpha=\tau_v/\tau_c$. In the left panel $\tau_c$ is fixed and $\tau_v$ is changed. In the right panel $\tau_v$ is fixed and $\tau_c$ is changed. } 
    \label{fig1}
\end{figure*}
\section{Physical significance of the quantum energy partition theorem} 
As we write in the Introduction, various expressions for the kinetic energy of a free Brownian particle can be found both in original papers and the well-known books, e.g. 
Eq. (83) in Ref. \cite{hakim}, Eq. (4.14) in Ref. \cite{ford88}, equation for the second moment of momentum in Table 2 of Ref. \cite{ingol1} and  Eq. (3.475) in Ref. \cite{breuer}.  
The form of $E_k$ can also be deduced from the fluctuation-dissipation relation obtained in the framework of the linear response theory which relates relaxation of a weakly perturbed system to the spontaneous fluctuations in thermal equilibrium, see e.g. Eq. (124.10) in Ref. \cite{landau}, Eq. (17.19g) in Ref. \cite{zubarev} and Eq. (3.499) in Ref. \cite{breuer}. 
All expressions for $E_k$ should be equivalent although they are written in different forms. However, our specific formula  (\ref{Ek}) allows to reveal a new face of the old problem and formulate new interpretations:\\
I. The mean kinetic energy $E_k$ of a free quantum particle equals the average kinetic energy $\langle \mathcal{E}_k \rangle$ of the thermostat degree of freedom, i.e. $E_k = \langle \mathcal{E}_k \rangle$. {\it Mutatis mutandis}, the form of this statement is exactly the same as for classical systems: The mean kinetic energy of a free classical  particle equals the average kinetic energy  of the thermostat degree of freedom.\\
II. The function $\mathbb{P}(\omega)$ is a probability density, i.e. it is non-negative and normalized  on the interval $(0, \infty)$. From the probability theory it follows that there exists a random variable $\xi$ for which $\mathbb{P}(\omega)$ is its probability distribution. Here, this random variable is interpreted as frequency of thermostat oscillators.\\
III.  Eq. (\ref{P}) can be converted to the transparent form
\begin{equation}
	\label{PF}
	\mathbb{P}(\omega) = \frac{2}{\pi} \, \int_0^{\infty}dt\, R(t) \cos(\omega t).   
\end{equation}
Thus the probability distribution $\mathbb{P}(\omega)$ is   a cosine Fourier  transform of the response function $R(t)$ which solves the generalized Langevin equation (\ref{GLE2}).\\
IV. Thermostat oscillators contribute to $E_k$ in a non-uniform way according to the probability distribution $\mathbb{P}(\omega)$. The form of this distribution depends on the response function in which full information on the thermostat modes and system-thermostat interaction is contained. \\
V. For high temperature,  Eq. (\ref{ho}) is approximated by  $\mathcal{E}_k(\omega) = k_B T/2$ and from Eq. (\ref{Ek}) we obtain the relation  $E_k =k_BT/2$, i.e. Eq. (\ref{Ek}) reduces to the energy equipartition theorem for classical systems.  

The next comment concerns the relation of Eq. (\ref{Ek}) with the fluctuation-dissipation theorem derived in the linear response theory. We adapt Eq. (124.10) from the Landau-Lifshitz book \cite{landau} in order to get the kinetic energy of the quantum particle, namely, 
\begin{equation}
	\label{landau}
	E_k= \frac{1}{2M} \langle p^2 \rangle = \frac{\hbar}{2\pi M}\int_0^\infty d\omega\, \coth{\left[\frac{\hbar\omega}{2k_BT}\right]} \,\alpha''(\omega),
\end{equation}
where $\alpha''(\omega)$ is the imaginary part of the generalized susceptibility $\alpha(\omega) = \alpha'(\omega) + i\alpha''(\omega)$. 
By direct comparison of Eq. (\ref{Ek}) and (\ref{landau})  we find the non-trivial relation between the probability distribution and the imaginary part of the generalized susceptibility, 
\begin{equation}
	\label{relation}
	\mathbb{P}(\omega) = \frac{2}{\pi} \frac{\alpha''(\omega)}{M \omega}.
\end{equation}
The second example is Eq. (4.14) in Ref. \cite{ford88}: 
\begin{equation}
	\label{ford}
	E_k = \frac{\hbar}{2\pi} \int_0^\infty d\omega \, \coth{\left[\frac{\hbar\omega}{2k_BT}\right]}\,M\omega^2 \,  \mbox{Im} [\alpha(\omega + i 0^+)],  
\end{equation}
where $\alpha(\omega)$ is also called susceptibility, which is not the same as in Eq. (\ref{landau}). Again, if we compare Eqs. (\ref{Ek}) and (\ref{ford}) then we can find the relation between $\mathbb{P}(\omega)$ and $\mbox{Im}[\alpha(\omega + i 0^+)]$. But now we get 
  \begin{equation}
	\label{relation2}
	\mathbb{P}(\omega) = \frac{2}{\pi} M\omega \, \mbox{Im} [\alpha(\omega + i 0^+)]. 
\end{equation}
We presented only two examples and to avoid confusion the reader should be careful with such relations because they depend on the specific form of the expression for $E_k$. Paraphrasing, "various authors present the same topic differently".

Overall, taking into account the non-trivial relation between the probability distribution $\mathbb{P}(\omega)$ and the imaginary part of the generalized susceptibility $\alpha''(\omega)$ we may say that our principle for quantum partition of energy (\ref{Ek})  can be seen as a specific form of the fluctuation-dissipation theorem of the Callen-Welton type although  it would be rather difficult to guess the form of $\mathbb{P}(\omega)$ knowing only the formula for fluctuation-dissipation theorem.  Finally, we note that Eq. (\ref{relation}) establishes the relation between the probability distribution $\mathbb{P}(\omega)$ and the generalized susceptibility $\alpha''(\omega)$. It means that features of the  quantum environment described by $\mathbb{P}(\omega)$ may be experimentally inferred from the measurement of the linear response of the system to an applied perturbation given as the corresponding classical susceptibility, e.g. electrical or magnetic. Consequently, according to our results the latter quantity may open a new pathway to study quantum open systems.

\section{Analysis of the probability distribution $ \mathbb{P}(\omega)$}
In the case of classical systems the averaged kinetic energy of the Brownian particle equals $E_k=k_BT/2$ and all thermostat oscillators have the same averaged kinetic energy $\mathcal E_k=k_BT/2$ which  does not depend on the frequency of a single oscillator. In the quantum case, $\mathcal E_k=\mathcal E_k(\omega)$  depends on the oscillator frequency $\omega$ and  oscillators of various frequencies contribute to $E_k$ with various probabilities. Therefore it is interesting to reveal which frequencies are more or less probable in dependence on the dissipation mechanism.  The impact of various dissipation mechanisms   can be analysed via one of three quantities: the dissipation kernel $\gamma(t)$, the correlation function $C(t)$ of the random force $\eta(t)$ or the spectral density $J(\omega)$. In our view, this mechanism can be  intuitively modelled by various forms of the damping kernel $\gamma(t)$.  Therefore in the following section we will examine  properties of the probability distribution $\mathbb P(\omega)$ for several classes of  $\gamma(t)$.
\begin{figure*}[t]
	\centering
	\includegraphics[width=0.39\linewidth]{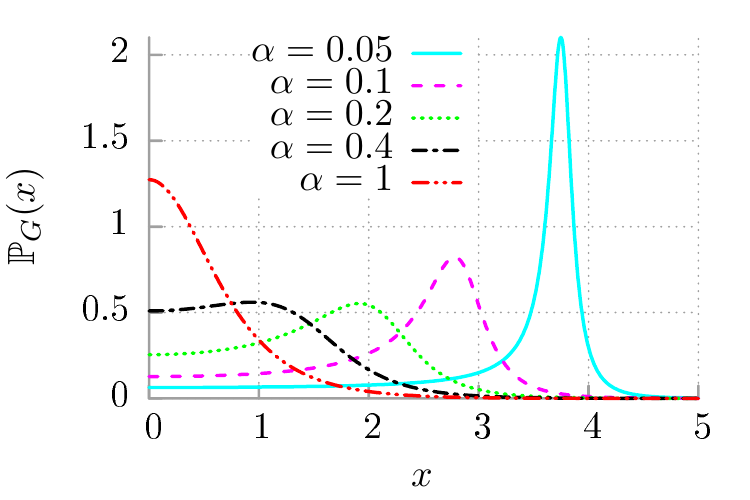}
	\includegraphics[width=0.39\linewidth]{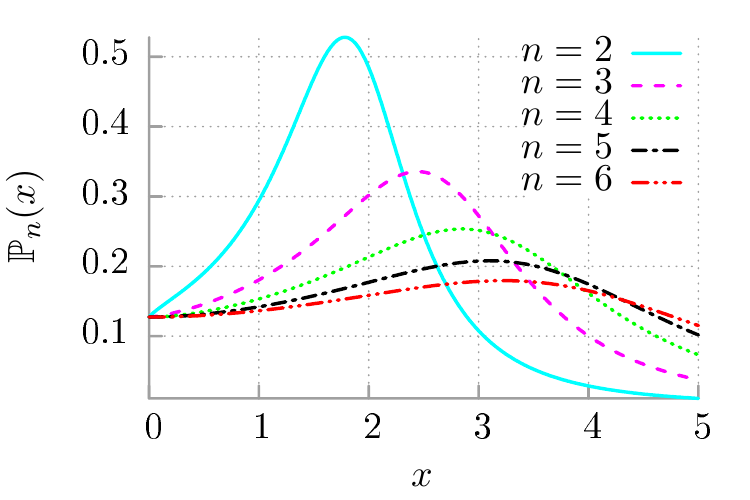}
		\caption{(color online): Left panel: The case of Gaussian decay of the memory kernel \mbox{$\gamma_G(t) = (\gamma_0/\sqrt{\pi} \tau_c) \, \mbox{e}^{-t^2/\tau_c^2}$}. Probability distribution $\mathbb{P}_G(x)$ is presented for different values of the dimensionless parameter $\alpha$ ($M$ and/or $\gamma_0$ is changed, $\tau_c$ is fixed). 
	 Right panel: The probability distribution $\mathbb{P}_{n}(x)$ is depicted for different values of the power exponent $n$ appearing in the generalized algebraic decay of the dissipation kernel $\gamma_{n}(t) = [(n-1)/2] \, \gamma_0 \tau_c^{n-1} /(t + \tau_c)^n$. The dimensionless parameter $\alpha = 0.1$.}
	\label{fig22}
\end{figure*}
\subsection{Drude model}
As the first step we assume the dissipation function $\gamma(t)$ to be in the form
\begin{equation}
	\label{g-drude}
	\gamma_D(t) = \frac{\gamma_0}{2 \tau_c}e^{-t/\tau_c}
\end{equation}
with two non-negative parameters $\gamma_0$ and $\tau_c$. The first one $\gamma_0$ is the particle-thermostat coupling strength and has the unit $[\gamma_0]=[kg/s]$, i.e. the same as the friction coefficient in the Stokes force. The second parameter $\tau_c$ characterizes time scale on which the system exhibits memory (non-Markovian) effects. Due to fluctuation-dissipation theorem $\tau_c$ can be also viewed as the primary correlation time of quantum thermal fluctuations. This exponential form of the memory function is known as a Drude model and it has been frequently considered in coloured noise problems. We choose the above scaling to ensure that if $\tau_c \to 0$  the function   $\gamma_D(t)$ is proportional to the Dirac delta  and the integral term in the generalized Langevin equation reduces to the frictional force of the Stokes form. Other damping kernels considered in the later part of this section also possesses this scaling property. With (\ref{diss}), instead of determining $\gamma(t)$, one can equivalently specify the spectral density of thermostat modes which for the Drude damping reads
\begin{equation}
	\label{j-drude}
	J_D(\omega) = \frac{1}{\pi}\frac{\gamma_0}{1 + \omega^2\tau_c^2}.
\end{equation}
From  Eq.   (\ref{Pp})  we get the following expression for the probability density 
\begin{equation} \label{P_D}
	\mathbb{P}(\omega) = \frac{1}{\pi} \, \frac{\mu_0\varepsilon^2(\omega^2 + \varepsilon^2)}{\omega^2 [\omega^2 + \varepsilon (\varepsilon -\mu_0/2)]^2 + \mu_0^2\varepsilon^4/4}, 
\end{equation}
where $\mu_0 = \gamma_0/M$ defines the rescaled coupling strength of the Brownian particle to the thermostat  and $\varepsilon = 1/\tau_c$ is the Drude frequency.  There are two control parameters 
$\varepsilon$ and $\mu_0$ which have the unit of frequency or equivalently two time scales: the memory time $\tau_c$ and $\tau_v = M/\gamma_0 = 1/\mu_0$ which in the case of a classical free Brownian particle defines the velocity relaxation time.  

If we want to analyse the impact of the particle mass $M$ or the coupling $\gamma_0$  we should use the following scaling 
\begin{equation} \label{scalx}
	x = \omega \tau_c =\frac{\omega}{\varepsilon},  
\end{equation}
which yields the expression
\begin{equation}
	\label{p-drude}
\mathbb P_D(x) = \varepsilon \mathbb{P}\left(\varepsilon x\right) = \frac{2}{\pi} \, \frac{2 \alpha(x^2 + 1)}{x^2[2\alpha (x^2 +1) -1]^2 + 1},
\end{equation}
where
\begin{equation} \label{alfa}
\alpha =  \frac{M}{\tau_c \gamma_0} =\frac{\varepsilon}{\mu_0} =\frac{\tau_v}{\tau_c}
\end{equation}
is the ratio of two characteristic times.   
It is remarkable that this probability distribution does not depend on these three parameters separately but only on one parameter $\alpha$ being their specific combination. We should remember that $\tau_c$ is fixed in this scaling.  In Fig. \ref{fig1} we present the probability distribution $\mathbb{P}_D(x)$ for different values of the parameter $\alpha$. We can observe that the thermostat oscillators contribute to the kinetic energy $E_k$ in a non-homogeneous way. There is the most probable value of $\mathbb P_D(x)$  indicating the optimal oscillator frequency $x_M$   which brings the greatest contribution to the kinetic energy of the Brownian particle. As it is illustrated in the panel,   $x_M$  is inversely proportional to $\alpha$: for small values of $\alpha$  mainly oscillators of high frequency contribute to $E_k$ whereas for large values of $\alpha$ primarily low frequencies. As $\alpha$ increases $x_M \to 0$  and $\mathbb P_D(x)$ becomes a monotonically decreasing function (not depicted).  In other words it means that e.g. when the coupling strength between the system and thermostat $\gamma_0$ is strong 
then contribution of high-frequency oscillators to $E_k$ is most pronounced; if the particle mass $M$ increases the optimal frequency $x_M$ decreases.

Next  we analyse the influence of the memory time $\tau_c$ on the probability distribution $\mathbb{P}(\omega)$. For this purpose we should  use another scaling:  
\begin{equation} \label{scaly}
	y = \frac{\omega}{\mu_0}.
\end{equation}
It leads to the expression
\begin{equation}
	\tilde{\mathbb{P}}_D(y) = \mu_0\mathbb{P}(\mu_0 y) = \frac{1}{\pi}\frac{\alpha^2(y^2 + \alpha^2)}{y^2 [y^2 + \alpha(\alpha - 1/2)]^2 + \alpha^4/4}, 
\end{equation}
with the same dimensionless parameter $\alpha$ defined in (\ref{alfa}).  In the right panel of Fig. \ref{fig1} we present this distribution for selected values of $\alpha$. It follows that for small values of the parameter $\alpha$, or equivalently for long memory time $\tau_c$, the distribution is notably peaked in the region of low frequency modes. Then it rapidly decreases to zero. Consequently only slowly vibrating thermostat oscillators contribute significantly to the kinetic energy of the particle. The situation is quite different for short memory time $\tau_c$ (large values of $\alpha$). Then the distribution is flattened meaning that much wider window of oscillators frequency contribute  to $E_k$ in a similar way.

In the remaining part of the paper, we present the probability distribution $\mathbb P(\omega)$ without any scaling. The reader can easily reproduce both scalings. For the  scaling as in Eq. (\ref{scalx}), one can put $\epsilon =1$ and rescale $\mu_0 \to \mu_0/\varepsilon$ to get the distribution $\mathbb P_i(x)$ (the index $i$ indicates the form of the memory function).  For the  scaling as in Eq. (\ref{scaly}), one can put $\mu_0 =1$ and rescale $\varepsilon \to \varepsilon/\mu_0$ to get the distribution $\mathbb P_i(y)$. 
In the first scaling, one can analyse the influence of the particle mass $M$ and the particle-thermostat coupling $\gamma_0$. In the second scaling - the memory time $\tau_c$.
\begin{figure}[t]
	\centering
	\includegraphics[width=0.88\linewidth]{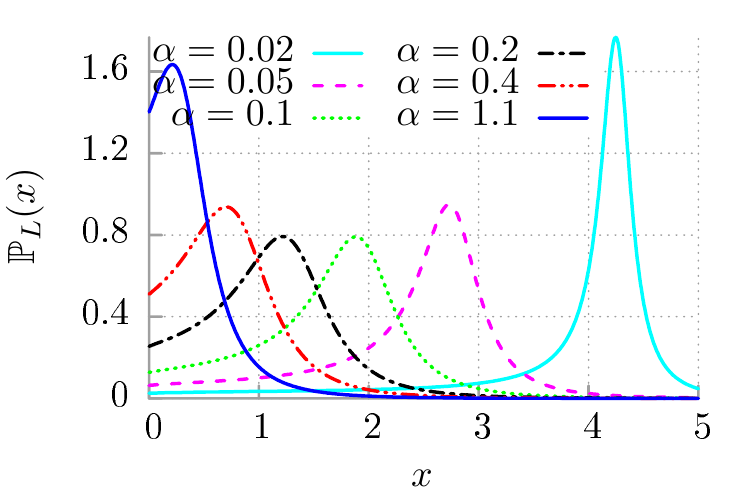}
	\caption{(color online): The probability distribution $\mathbb{P}_L(x)$ is depicted for the Lorentzian dissipation kernel \mbox{$\gamma_L(t) = \gamma_0 \tau_c / \pi(t^2 + \tau_c^2)$} and selected  values of the dimensionless parameter $\alpha$ ($M$ or $\gamma_0$ can be changed and $\tau_c$ is fixed).}
	\label{fig3}
\end{figure}
\ \ \\
\subsection{Gaussian decay}
Another possible choice of the dissipation kernel $\gamma(t)$ is the rapidly  decreasing  Gaussian function, namely, 
\begin{equation}
	\label{g-gauss}
	\gamma_G(t) = \frac{\gamma_0}{\sqrt{\pi} \tau_c}\, \mbox{e}^{-t^2/\tau_c^2}
\end{equation}
for which the corresponding spectral density is also Gaussian and reads
\begin{equation}
	\label{j-gauss}
	J_G(\omega) = \frac{\gamma_0}{\pi}\, \mbox{e}^{-\omega^2\tau_c^2/4}.
\end{equation}
In order to have an identical notation as in the previous case, we present  the probability distribution in the form ($\varepsilon =1/\tau_c$) 
\begin{widetext}
\begin{eqnarray}
	\label{p-gauss}
	\mathbb{P}_G(\omega) = \frac{4}{\pi \mu_0} \, \frac{ e^{-(\omega/ 4 \varepsilon)^2}}
	{\left[2 \omega/\mu_0   + i e^{-(\omega/ 4 \varepsilon)^2}  \mbox{Erf}\left(-i\omega/ 2 \varepsilon\right)\right]\left[ 2 \omega/\mu_0  - i  e^{-(\omega/ 4 \varepsilon)^2}  \mbox{Erf}\left(i\omega/ 2 \varepsilon \right) \right]},
\end{eqnarray}
\end{widetext}
where $\mbox{Erf}(z)$ is the error function 
\begin{equation}
	\mbox{Erf}(z) = \frac{2}{\sqrt{\pi}} \int_0^z dt\, e^{-t^2}.
\end{equation}
In Fig. \ref{fig22} we present this probability distribution $\mathbb{P}_G(x)$ [in the scaling (\ref{scalx})] for selected  values of $\alpha$ ($\tau_v = M/\gamma_0$ is changed and $\tau_c$ is fixed) . Similarly as in the case of the Drude model, the oscillator frequency $x_M$ which brings the greatest contribution to the kinetic energy of the particle is inversely proportional to the parameter $\alpha$. However, here we observe two differences: (i) at some interval of $\alpha$ the maximum of $\mathbb P_G(x)$ decreases as $\alpha$ increases and (ii) the half-width of  $\mathbb{P}_G(x)$ increases as $\alpha$ increases while for the Drude model it is almost  constant in a wide interval of $\alpha$. In this case, the impact of the memory time $\tau_c$  is similar to that as for the Drude dissipation, see the right panel of Fig. 1.
\begin{figure*}[t]
	\centering
	\includegraphics[width=0.39\linewidth]{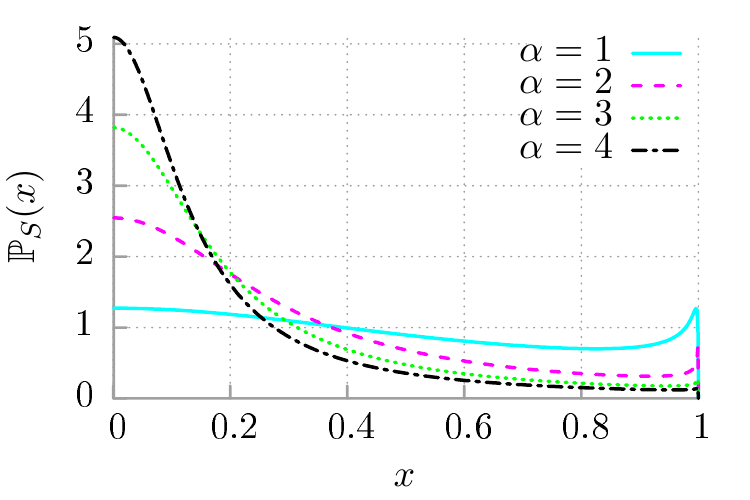}
	\includegraphics[width=0.39\linewidth]{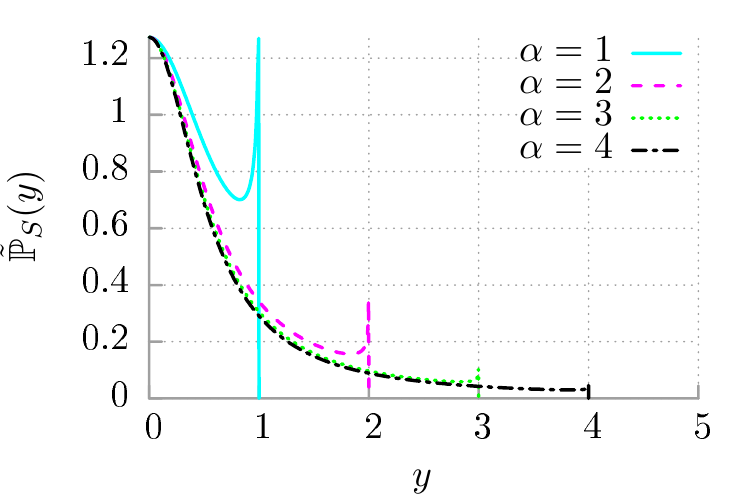}
	\caption{(color online): The probability distribution $\mathbb{P}_S(x)$ is presented for the oscillatory decay $\gamma_S(t) = (\gamma_0/\pi)\sin(t/\tau_c)/t$ (the Debye type model) and selected values of the dimensionless parameter $\alpha=\tau_v/\tau_c$.  In the left panel $\tau_c$ is fixed and $\tau_v=M/\gamma_0$ is changed. In the right panel $\tau_v$ is fixed and $\tau_c$ is changed.} 
	\label{fig4}
\end{figure*} 
\subsection{n-Algebraic decay}
Apart from two exponential forms  of the memory functions which we presented above one could model the dissipation function $\gamma(t)$ with  algebraic decay.  It is worth noting that the power-law decay of the  memory functions has been considered as a model of anomalous transport processes \cite{anomal1,anomal2}.
Here, we consider the class of functions 
\begin{equation}
	\label{g-algebraic-n}
	\gamma_{n}(t) = \frac{n-1}{2}\frac{\gamma_0 \tau_c^{n-1} }{(t + \tau_c)^n},
\end{equation}
where  $n \in \mathbb{N}$ and $n \geq 2$.  It has the same limiting Dirac delta form for 
$\tau_c\to 0$ as in two previous cases. 
The corresponding spectral density reads
\begin{equation}
	\label{j-algebraic-n}
	J_{n}(\omega) = \frac{(n-1)\gamma_0}{2\pi} \, \left[e^{-i \omega \tau_c} E_n(-i\omega \tau_c) + e^{i \omega \tau_c}E_n(i \omega \tau_c)\right] 
\end{equation}
and  $E_n(z)$ is the exponential integral, 
\begin{equation}
E_n(z) = \int_1^\infty dt\, \frac{e^{-zt}}{t^n}.
\end{equation}
The probability distribution takes the form 
\begin{widetext}
\begin{eqnarray}
\label{p-algebraic-n}
\mathbb{P}_{n}(\omega) = \frac{2(n-1)}{\pi \mu_0} 
  \frac{e^{-i\omega/ \varepsilon}E_n(-i \omega/ \varepsilon) + e^{i \omega / \varepsilon}E_n(i \omega/ \varepsilon)}{\left[(n-1)e^{-i \omega/ \varepsilon}E_n(-i \omega/ \varepsilon) - 2i \omega/\mu_0 \right]\left[(n-1)e^{i \omega/ \varepsilon}E_n(i \omega/ \varepsilon) + 2i \omega/\mu_0  \right]}. 
\end{eqnarray}
\end{widetext}
In Fig. \ref{fig22} we present the influence of the power exponent $n$ appearing in the dissipation function  $\gamma_{n}(t)$ on the probability distribution $\mathbb{P}_{n}(x)$ for fixed $\alpha = 0.1$. The conclusion is: an increase of the exponent $n$ causes progressive flattening of the probability density function. In other words, if the memory function decreases faster and faster to zero the wider spectrum of frequencies  of  the thermostat oscillators contribute to $E_k$.
\subsection{Lorentzian decay}
It  is interesting to compare the algebraic case for $n=2$ with the Lorentzian memory function  which reads
\begin{equation}
	\label{g-cauchy}
	\gamma_L(t) = \frac{\gamma_0}{\pi}\, \frac{\tau_c}{t^2 + \tau_c^2}.
\end{equation}
In the probability theory it is termed as the Cauchy distribution. Alternatively, it may be imposed by the following spectral density of thermostat modes,  
\begin{equation}
	\label{j-cauchy}
	J_L(\omega) = \frac{\gamma_0}{\pi}e^{-\omega \tau_c}.	
\end{equation}
Such a choice of the dissipation kernel leads  to the following probability distribution  ($\varepsilon =1/\tau_c$)
\begin{equation}
	\label{p-cauchy0}
	\mathbb{P}_L(\omega) = \frac{4\pi}{\mu_0} \, 
	\frac{e^{- \omega/\varepsilon}}{\pi^2 e^{-2\omega/\varepsilon} + c^2(\omega)},
\end{equation}
where 
\begin{eqnarray}
	c(\omega) = e^{-\omega/\varepsilon} \mbox{Ei}(\omega/\varepsilon) - e^{\omega/\varepsilon} \mbox{Ei}(-\omega/\varepsilon) -\frac{2\pi}{\mu_0} \omega 
\end{eqnarray}
and $\mbox{Ei}(z)$ is the exponential  integral defined as
\begin{equation}
\mbox{Ei}(z) = \int_{-\infty}^z \frac{e^{t}}{t} \, dt.
\end{equation}
We illustrate this probability distribution in Fig. \ref{fig3} for different values of the dimensionless parameter $\alpha = M/\tau_c \gamma_0$. The oscillator frequency $x$ which brings the greatest contribution to the kinetic energy of the particle is inversely proportional to the parameter $\alpha$. Again,  as it was in the previous cases, the magnitude of the maxima in the probability distribution $\mathbb{P}_L(x)$ also depends on $\alpha$. For very small values of $\alpha$ one can note that high frequency modes almost exclusively contribute to the kinetic energy of the particle. 
\begin{figure}[t]
	\centering
	\includegraphics[width=0.88\linewidth]{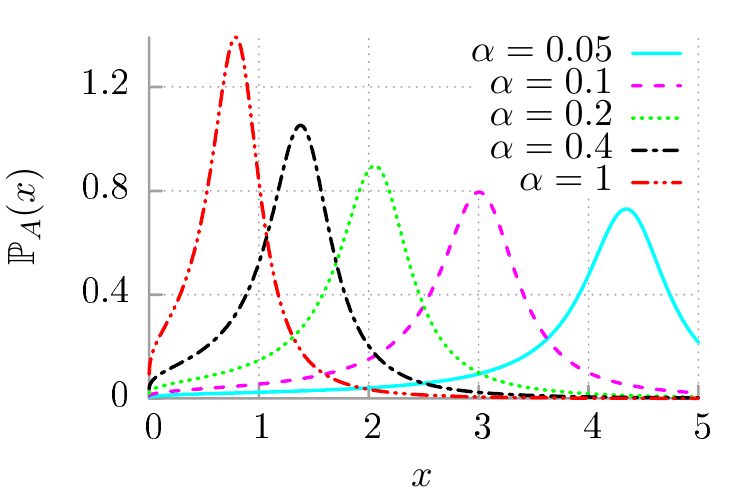}
	\caption{(color online): Algebraic decay of the dissipation function \mbox{$\gamma_A(t) = \gamma_0/(t + \tau_c)$}. Probability distribution $\mathbb{P}_A(x)$ is presented for different values of the dimensionless parameter $\alpha$. }
	\label{fig5}
\end{figure}
\subsection{Debye type model: Algebraically decaying oscillations}
The next  example of this series is the oscillatory memory function \cite{zwan} 
\begin{equation}
	\label{g-sin}
	\gamma_S(t) = \frac{\gamma_0}{\pi}\frac{\sin{(t/\tau_c)}}{t}. 
\end{equation}
which takes both positive and negative values. One can show that, via the fluctuation-dissipation relation,  the quantum noise $\eta(t)$ exhibits anti-correlations.  
The spectral density is of the Debye type \cite{zwan} 
\begin{equation}
	\label{j-sin}
	J_S(\omega) = \frac{\gamma_0}{\pi} \, \theta\left(\frac{1}{\tau_c} - \omega\right), 
\end{equation}
where $\theta(x)$ denotes the Heaviside step function. This spectral density  
is constant $J(\omega)=\gamma_0/\pi$ on the  {\it compact} support $[0,1/\tau_c]$ determined by the memory  time $\tau_c$ or the cut-off frequency $\varepsilon = 1/\tau_c$.  Under this assumption the probability density $\mathbb{P}_S(\omega)$ reads
\begin{widetext}
\begin{equation}
	\label{p-sin}
	\mathbb{P}_S(\omega) = \frac{4\pi}{\mu_0} \, \frac{\theta(\varepsilon-\omega )}{\pi^2(1+ 4\omega^2/\mu_0^2) + 4\mbox{arctanh}(\omega/\varepsilon )[\mbox{arctanh}(\omega/\varepsilon ) - 2\pi \omega/\mu_0]}
\end{equation}
\end{widetext}
and has the same support as $J(\omega)$ in the interval $[0,\varepsilon]$. In Fig. \ref{fig4} we present the probability density $\mathbb{P}_S(x)$ for selected values of the dimensionless parameter $\alpha$ in two various scalings. In the left panel, the memory time is fixed and the coupling $\gamma_0$ or the mass $M$ is changed. Again, when e.g. $\gamma_0$ decreases (i.e. $\alpha$ increases)  more and more oscillators of low frequency contribute to $E_k$.
\begin{figure*}[t]
	\centering
	\includegraphics[width=0.39\linewidth]{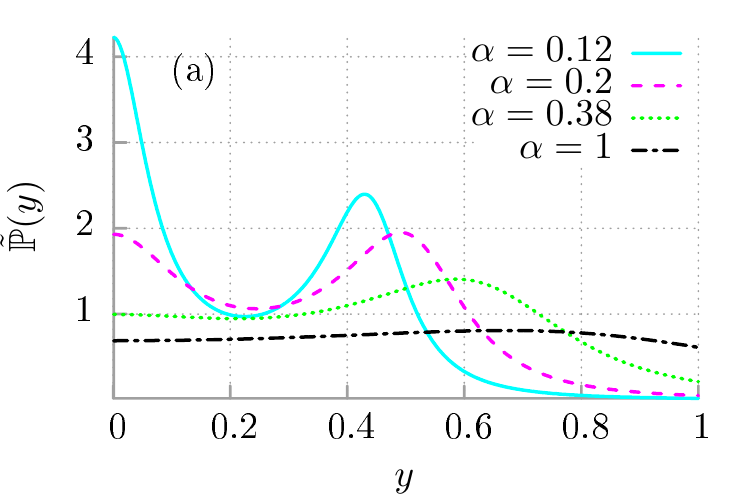}
	\includegraphics[width=0.39\linewidth]{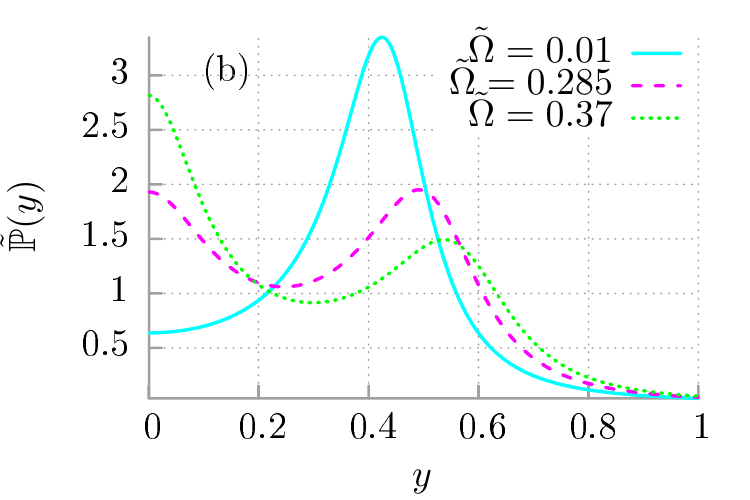}
	\caption{(color online): Panel (a): The probability distribution $\tilde{\mathbb{P}}(y)$ scaled according to Equation (\ref{scaly}) is depicted for exponentially decaying oscillations with $\gamma_E(t) = (\gamma_0/\tau_c) e^{-t/\tau_c}\cos{(\Omega t)}$ and different values of  $\alpha = \varepsilon/\mu_0$ and fixed  $\tilde{\Omega} = \Omega/\mu_0 = 0.285$. Panel (b): The same  $\tilde{\mathbb{P}}(y)$  is presented for selected dimensionless frequencies $\tilde{\Omega}$ of the memory function and fixed $\alpha = 0.2$.}
	\label{fig6}
\end{figure*}
\subsection{Slow algebraic decay}
In this subsection we consider slow algebraic decay of the memory kernel assuming  
\begin{equation}
	\label{g-algebraic}
	\gamma_A(t) = \frac{\gamma_0}{t+\tau_c}. 
\end{equation}
This dissipation function does not tend to the Dirac delta when $\tau_c \to 0$ (the limit does not exist at all) and therefore is not placed in the subsection IVC.  The corresponding spectral density has the form
\begin{equation}
	\label{j-algebraic}
	J_A(\omega) = \frac{2\gamma_0}{\pi} \, a(\omega).  
\end{equation}
The probability distribution reads  
\begin{equation}
	\label{p-algebraic3}
	\mathbb{P}_A(\omega) = \frac{2}{\pi \mu_0} \, \frac{a(\omega)}{a^2(\omega) + 
	[b(\omega) -\omega/\mu_0]^2}, 
	\end{equation}
where ($\varepsilon =1/\tau_c$)
\begin{eqnarray}
	&a(\omega)  = -\mbox{ci}(\omega/\varepsilon) \cos(\omega/\varepsilon) -\mbox{si}(\omega/\varepsilon)\sin(\omega/\varepsilon), \\
&b(\omega)  = \mbox{ci}(\omega/\varepsilon) \sin(\omega/\varepsilon) -\mbox{si}(\omega/\varepsilon)\cos(\omega/\varepsilon). 	
	\end{eqnarray}
The functions $\mbox{ci}(z)$ and $\mbox{si}(z)$ are cosine and sine integrals defined as 
\begin{eqnarray}
	\mbox{ci}(z) &=& -\int_z^{\infty}  \frac{\cos{t}}{t} \, dt, \\
	\mbox{si}(z) &=& -\int_z^{\infty} \frac{\sin{t}}{t} \, dt.
\end{eqnarray}
In Fig. \ref{fig5} we depict $\mathbb P_A(x)$  for different values of the dimensionless parameter $\alpha$. The same as before, the optimal frequency of oscillator which has the largest impact on the kinetic energy is inversely proportional to $\alpha$. Qualitatively, it looks similar to the case of the Drude model, c.f. Fig. \ref{fig1}. However, only for large value of $\alpha$ contribution of harmonic modes of lowest frequency $x \to 0$ differs significantly from zero.

Overall, the common characteristic feature of all cases presented above is that the probability distribution $\mathbb{P}(x)$ occurring in the quantum law for energy equipartition depends only on one dimensionless parameter $\alpha = M/\tau_c \gamma_0$. Moreover, for a small  value of this parameter (the strong particle-thermostat coupling) one typically finds the bell-shaped probability density with a pronounced maximum for high frequency $x_M$ which is inversely proportional to the magnitude of $\alpha$. 
For large value of $\alpha$, thermostat oscillators of low frequency dominate in contribution to the kinetic energy of the Brownian particle.
\begin{figure*}[t]
	\centering
	\includegraphics[width=0.39\linewidth]{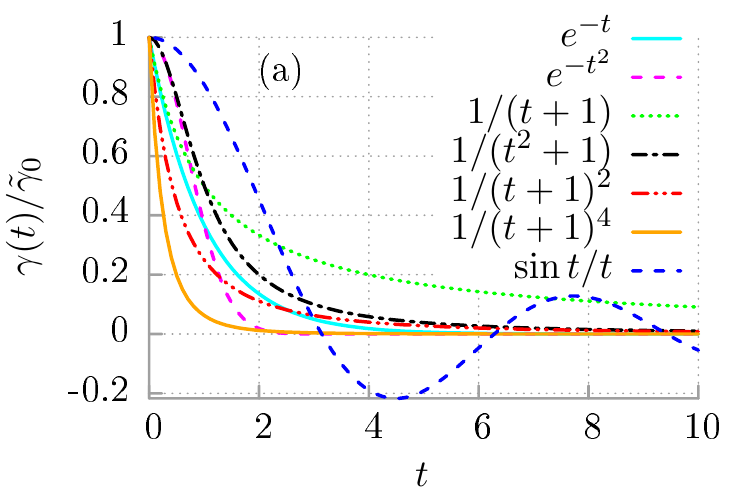} 
	\includegraphics[width=0.39\linewidth]{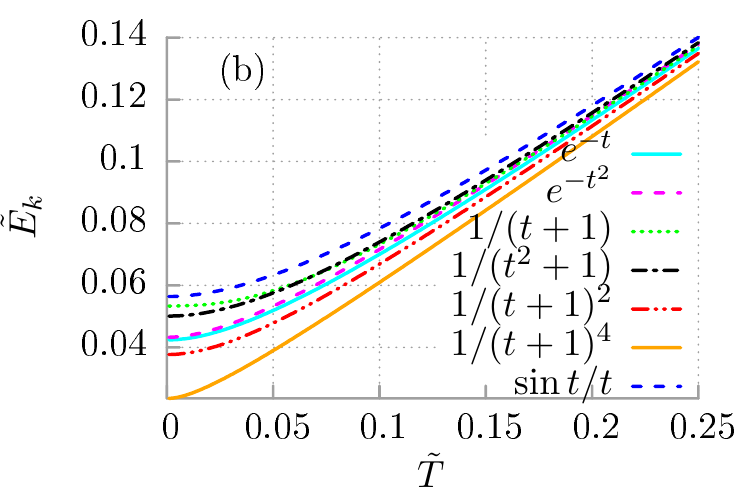} \\
	\includegraphics[width=0.39\linewidth]{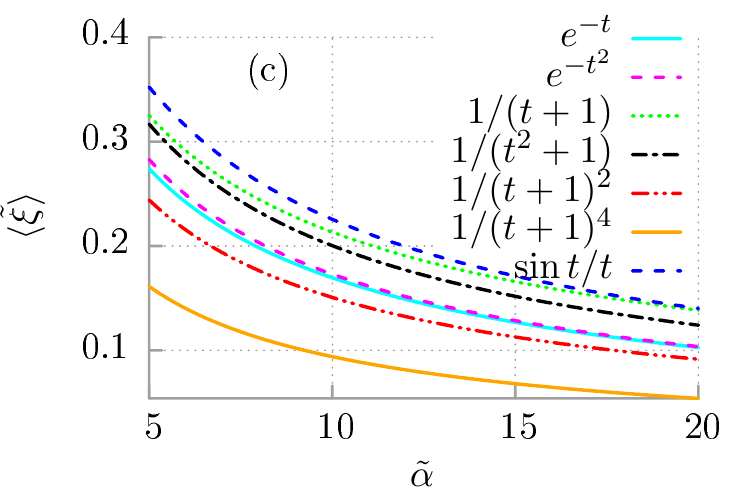}
	\includegraphics[width=0.39\linewidth]{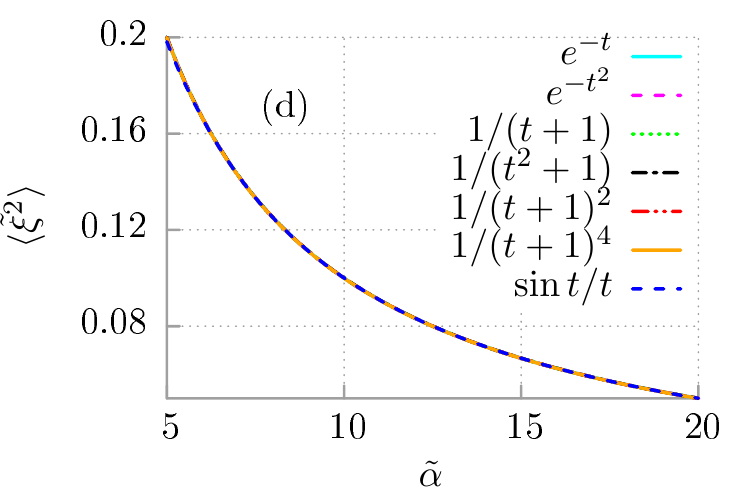}
	\caption{(color online): Panel (a): The normalized memory functions $\gamma(t)/\tilde\gamma_0$ representing various dissipation mechanisms. Panel (b): The dimensionless kinetic energy $\tilde E_k= \tau_c {E}_k/\hbar$ of the free Brownian particle presented versus dimensionless temperature $\tilde T = \tau_c k_B T/\hbar$ and various forms of  $\gamma(t)$. Panel (c): The first moment $\langle \tilde\xi \rangle = \tau_c \langle \xi \rangle$ and  panel (d): the second moment $\langle \tilde\xi^2 \rangle = \tau_c^2\langle \xi^2 \rangle$ depicted versus the dimensionless parameter $\tilde\alpha = M/\tilde\gamma_0 \tau_c^2$ for different variants of the damping kernel $\gamma(t)$.}  
	\label{fig7}
\end{figure*}
\subsection{Exponentially decaying oscillations}
As the last example, we consider a generalization of the Drude model in the form of exponentially decaying oscillations \cite{bialas}, 
\begin{equation}
	\label{g-entropy}
	\gamma_E(t) = \frac{\gamma_0}{\tau_c}e^{-t/\tau_c}\cos{(\Omega t)},
\end{equation}
where in addition to the previously defined parameters $\gamma_0$ and $\tau_c$,  now $\Omega$ is the frequency in the relaxation process of the particle momentum. Also in this case, the quantum noise $\eta(t)$ exhibits anti-correlations.  The limiting case $\Omega = 0$ corresponds to the Drude model of dissipation. Such a choice of the damping kernel leads to the following spectral density
\begin{equation}
	\label{j-entropy}
	J_E(\omega) = \frac{2}{\pi}\frac{\gamma_0\varepsilon^2(\varepsilon^2 + \omega^2 + \Omega^2)}{(\varepsilon^2 + \omega^2)^2 + 2\Omega^2(\varepsilon^2 - \omega^2) + \Omega^2},
\end{equation}
where $\varepsilon = 1/\tau_c$. From  the quantum law for the partition of energy we obtain the probability distribution in the form \cite{bialas} 
\begin{equation}
	\label{p-entropy}
	\mathbb{P}(\omega)  = \frac{2}{\pi} \, \frac{ \mu_0 \varepsilon^2 \left( \omega^{2} + \varepsilon^{2} + \Omega^{2} \right) } 
	{ \omega^{2}\left[(\omega^2+  \varepsilon^{2} -  \Omega^{2}  -\mu_0 \varepsilon)^2 
	+ 4 \varepsilon^2\Omega^{2}\right] +  \mu_0^2 \varepsilon^4}. 
\end{equation}
The parameter $\mu_0 = \gamma_0/M$ defines the rescaled coupling strength of the Brownian particle to thermostat. We note that in the considered case there are three characteristic frequencies $\mu_0$, $\varepsilon$ and $\Omega$ or equivalently three time scales which are equal to the reciprocals of these frequencies. This observation must be contrasted with all previously considered damping kernels leading to two characteristic time scales. The kinetic energy of the free Brownian particle with the exponentially decaying oscillations in the dissipation function was analysed in detail in Ref. \cite{bialas}. Instead, here we focus on the properties of the probability density occurring in the quantum energy partition theorem. The influence of the coupling strength $\mu_0$ on  $\mathbb{P}(\omega)$ is similar to that of the Drude model: there is only one maximum for a fixed value of the coupling strength $\mu_0$. For larger values of the latter it is shifted to the right indicating that oscillators of the higher frequency bring the greatest contribution to the kinetic energy of the particle.

The influence of the reciprocal of the correlation time $\varepsilon = 1/\tau_c$ is depicted in Fig. \ref{fig6}(a). In this case, we scale Eq.  (\ref{p-entropy}) as in  (\ref{scaly}), namely $y = \omega/\mu_0$. The dimensionless parameters are $\alpha = \varepsilon/\mu_0$ and $\tilde{\Omega} = \Omega/\mu_0$.  Due to the interplay of two characteristic time scales associated with the parameters $\alpha$ and $\tilde{\Omega}$ we observe here qualitatively new features. For large values of $\alpha \gg \tilde{\Omega}$ the distribution is almost flat indicating that all oscillators of thermostat contribute equally to the kinetic energy of the system. When the the characteristic frequency $\alpha$ is slightly larger than the other one $\alpha > \tilde{\Omega}$ a single maximum is born. When the opposite situation occurs, i.e. $\alpha < \tilde{\Omega}$ then the distribution $\tilde{\mathbb{P}}(y)$ exhibits a clear bimodal character. It means that both oscillators of low and moderate frequency play important role. Further decrease of $\alpha$ extinguishes the contribution of higher frequencies at the favour of the near zero frequency modes which are then the most pronounced ones. 

Last but not least, we elaborate on the impact of the oscillation frequency $\Omega$. We keep the scaling with respect to the system-thermostat coupling strength $\mu_0$. In Fig. \ref{fig6}(b) we present the probability distribution $\tilde{\mathbb{P}}(y)$ for a few values of the dimensionless frequency $\tilde{\Omega} = \Omega/\mu_0$ and  fixed $\alpha = \varepsilon/\tau_c = 0.2$. The result confirms our earlier observation that due to interplay of two characteristic time scales the probability density may be bimodal. It is realized when the magnitude of $\tilde{\Omega}$ and $\alpha$ is comparable. For very small $\tilde{\Omega}$ the distribution $\tilde{\mathbb{P}}(y)$ possesses one very pronounced maximum, whereas for large $\tilde{\Omega}$ it becomes a monotonically decreasing function of the dimensionless frequency $y$.
\section{Statistical moments of the probability distribution $\mathbb{P}(\omega)$} 
Let us now discuss statistical moments of the random variable $\xi$ distributed according to the probability density $\mathbb{P}(\omega)$, 
\begin{equation}
	\langle \xi^n \rangle = \int_0^\infty d\omega \, \omega^n \mathbb{P}(\omega).
\end{equation}
A caution is needed since not all moments may exist, e.g. for  the distribution 
(\ref{P_D}). The first two of them have a clear physical interpretation \cite{arxiv2018}. The first moment, i.e. the mean value $\langle \xi \rangle$ of the random variable $\xi$ is proportional to the kinetic energy $E_k$ of the Brownian particle at zero temperature 
$T = 0$, namely, 
\begin{equation}
	E_0 = E_k(T = 0) = \frac{\hbar}{4}  \, \langle \xi \rangle. 
\end{equation}
 The second moment $\langle \xi^2 \rangle$ is proportional to the first correction of kinetic energy $E_k$ in the high temperature regime, 
\begin{equation}
	E_k = \frac{1}{2} k_B T + \frac{\hbar^2}{24 k_B T } \, \langle \xi^2 \rangle.
\end{equation}
We  note that averaged kinetic energy $E_0$ at zero temperature $T=0$ is non-zero  for all values of the system parameters. It is so because of intrinsic quantum vacuum fluctuations. Moreover, $E_k$   monotonically increases from some non-zero value to infinity when temperature goes to infinity. If we want to compare impact of various dissipation mechanisms on $E_k$ we have to change the scaling of all dissipation functions $\gamma(t)$. Now, we re-define $\gamma(t)$ in such a way that for all memory functions  $\gamma(0)=\tilde\gamma_0$, where $\tilde\gamma_0$ still characterizes the particle-thermostat coupling but now it has the unit $[\tilde\gamma_0]=[kg/s^2]$. 
E.g. for the Drude model $\gamma_D(t) = \tilde\gamma_0 \,\mbox{exp}(-t/\tau_c)$ or  for the Lorenzian shape $\gamma_L(t) = \tilde\gamma_0 /[(t/\tau_c)^2+1]$, see panel (a) of Fig. \ref{fig7}, where all $\gamma(t)$ assume the same value for $t=0$.  In the classical case, it would correspond to the  
fixing of the second moment of the random force $\eta(t)$. In Section 6, we define 
$\gamma(t)$ in such a way that  $\gamma(t)$ tends to the Dirac delta when the memory time $\tau_c \to 0$, which in the classical case corresponds to Gaussian  white noise of the random force $\eta(t)$.  

In Fig.  \ref{fig7}(b) we compare the kinetic energy $E_k$ for different forms of the memory function $\gamma(t)$.  The various  curves  $E_k$ versus temperature  never intersect each other for the same set of parameters. Therefore  it is sufficient to analyse the energy only at zero temperature $E_0 \propto \langle \xi \rangle$. We present this characteristic in Fig. \ref{fig7}(c) where we depict the dimensionless first moment $\langle \tilde\xi \rangle = \tau_c \langle \xi \rangle$   of the probability density $\mathbb{P}(\omega)$ versus the dimensionless parameter  $\tilde\alpha = M/\tilde\gamma_0 \tau_c^2$. In calculations we scale $\omega =x/\tau_c$ like in (\ref{scalx}) with fixed $\tau_c$. First, we note that in all cases the averaged kinetic energy at zero temperature decreases when the parameter $\tilde\alpha$ increases. We recall that it translates to either (i) increase of the particle mass $M$ or (ii) decrease  of the coupling strength $\tilde\gamma_0$. Moreover, we can see that for the n-algebraic decay ($n=4$ for green and $n=2$ for red curves, respectively) the kinetic energy at zero temperature $E_0$ is smaller than for other memory functions. 
The negligible difference is observed for the Drude and Gaussian decay. The largest kinetic energy is induced by the Debye type dissipation. In the high temperature regime (panel (d) of  Figure \ref{fig7}),  the correction $\langle \tilde\xi^2 \rangle = \tau_c^2 \langle \xi^2 \rangle$ depends very weakly on the form of $\gamma(t)$ and the differences are indistinguishable. Finally, at $T=0$, the energy $E_0$ increases starting from zero for $\tau_c\to 0$ and  saturates to  a finite value as $\tau_c$ is longer and longer (not depicted).   
\section{Summary}
In this work we have revisited an archetype model of quantum Brownian motion formulated in terms of the generalized quantum Langevin equation for a free particle interacting with a large number of independent oscillators that form thermal reservoir. In particular, we analysed the impact of various dissipation mechanisms on the averaged kinetic energy $E_k$ of the Brownian particle. For this purpose we harvested the recently formulated quantum law for partition of energy. It expresses the kinetic energy $E_k$ of the particle as the mean kinetic energy per one degree of freedom of the thermostat oscillators $E_k = \langle \mathcal{E}_k \rangle$. The averaging over frequencies $\omega$ of those oscillators is performed according to the probability distribution $\mathbb{P}(\omega)$ which is related to the dissipation kernel $\gamma(t)$ via the quantum partition theorem. We focused mainly on the influence of the form of the dissipation function on characteristic features of the probability density $\mathbb{P}(\omega)$.

We analysed multitude of dissipation mechanisms which are  grouped into two classes of the algebraic and exponential decay. Within each of them we considered the monotonic as well as oscillating decay. For the dissipation functions possessing two characteristic time scales associated with the relaxation time of the particle momentum $M/\gamma_0$ and the correlation time of quantum thermal fluctuations $\tau_c$ typically we observed the bell shaped probability distribution $\mathbb{P}(\omega)$. It means that there is an optimal oscillator frequency which brings the greatest contribution to the kinetic energy of the particle. The magnitude of this optimum is inversely proportional to the system-thermostat coupling strength $\gamma_0$. For large values of the latter the contribution of high frequency oscillators is most pronounced. We studied also impact of the memory  time $\tau_c$ on the shape of the distribution $\mathbb{P}(\omega)$. For long memory time $\tau_c$ the probability density is noticeable peaked whereas for short  $\tau_c$ the distribution is almost flat. Consequently, a decrease  of the memory time $\tau_c$ causes flattening of the probability density $\mathbb{P}(\omega)$. In this class of dissipation functions we have considered a peculiar case of the algebraically decaying oscillations $\gamma(t) \propto \sin{t}/t$. This choice leads to the distribution $\mathbb{P}(\omega)$ possessing a finite cut-off frequency which curiously depends on the correlation time of quantum fluctuations $\tau_c$. For dissipation mechanism with additional characteristic time scale associated with the period of oscillations $2\pi/\Omega$ qualitatively new features emerge in the density $\mathbb{P}(\omega)$. We exemplify this observation for the case of exponentially decaying oscillations. When the magnitude of $\tau_c$ and $2\pi/\Omega$ are similar then the probability distribution displays the bimodal character. This means that there are two characteristic frequencies of the thermostat oscillators which brings the significant contribution to the kinetic energy of the system.

We have demonstrated  that the quantum law for  energy partition in the present formulation is conceptually simple yet very powerful tool for analysis of quantum open systems. We hope that our work will stimulate its further successful applications.
\section*{Acknowledgement}
J. S. was supported by the Foundation for Polish Science (FNP) START fellowship and the Grant NCN 2017/26/D/ST2/00543. P. B. and J. {\L}. were supported by the Grant NCN 2015/19/B/ST2/02856.
\appendix
\section{Solution of the Langevin equation (\ref{GLE2})}
Eq. (\ref{GLE2}) is a linear integro-differential equation for the momentum operator $p(t)$.  
Because its integral part is a convolution, it can be solved by the Laplace transform method yielding 
\begin{equation}\label{pL} 
z \hat p_L(z) -p(0) + \frac{1}{M} \hat \gamma_L(z) \hat p_L(z) = - \hat \gamma_L(z) x(0) + \hat \eta_L(z),   
\end{equation}
where $\hat p_L(z)$,  $\hat \gamma_L(z)$ and $ \hat \eta_L(z)$ are the Laplace transforms of $p(t), \gamma(t)$ and $\eta(t)$, respectively (see  Eq. (\ref{fL}). The operators $p(0)$ and $x(0)$ are the momentum and coordinate operators of the Brownian particle at time $t=0$. 
From this equation it follows that  
\begin{equation}\label{pLL} 
\hat p_L(z) = \hat{R}_L(z) p(0) - \hat{R}_L(z)\hat \gamma_L(z) x(0) + \hat{R}_L(z)\hat \eta_L(z), 
\end{equation}
where
\begin{equation}\label{RL2} 
\hat{R}_L(z) = \frac{M}{Mz + \hat \gamma_L(z)}. 
\end{equation}  
 The inverse Laplace transform of (\ref{pLL}) gives the solution $p(t)$ for the momentum of the Brownian particle, namely, 
\begin{align}\label{p2(t)} 
p(t) &= R(t)p(0) - \int_0^t du\; R(t-u) \gamma(u)x(0)\nonumber \\  &+ \int_0^t du\; R(t-u) \eta(u),   
\end{align}
where the response function $R(t)$ is the inverse Laplace transform of the function $\hat{R}_L(z)$ in Eq. (\ref{RL2}).  
Because statistical properties of thermal noise $\eta(t)$ are specified, all statistical characteristics of the particle momentum $p(t)$ can be calculated, in particular its kinetic energy.  
\section{Kinetic energy in an equilibrium state}
In order to derive the averaged kinetic energy of the Brownian particle in the equilibrium state, we first calculate the symmetrized momentum-momentum correlation function $\symkowm{p(t)}{p(s)}$. For long times, $t\gg 1, s\gg 1$,   only the  last term of (\ref{p2(t)}) contributes and then   
\begin{widetext}
\begin{equation} \label{pp}
\symkowm{p(t)}{p(s)} = \int_0^{t} dt_1 \int_0^{s} dt_2 \; R(t-t_1) R(s-t_2) \symkowm{\eta(t_1)}{\eta(t_2)}. 
\end{equation}
\end{widetext}
Now, we express the correlation function $C(t_1-t_2)= \symkowm{\eta(t_1)}{\eta(t_2)}$ of quantum thermal noise by its Fourier transform, see Eq. (\ref{C}) in Appendix C, 
\begin{widetext}
\begin{eqnarray} \label{pTpS}
\symkowm{p(t)}{p(s)} = \int_0^{\infty} d\omega \; \hat{C}_F(\omega) \int_0^{t} dt_1 \int_0^{s} dt_2 \; R(t-t_1)  R(s-t_2) \cos \left[ \omega \left(t_1-t_2 \right) \right]. 
\end{eqnarray}
\end{widetext}
In particular, for $t=s$, it is the second statistical moment of the momentum, 
\begin{widetext}
\begin{equation} \label{pTpT}
\langle p^2(t)\rangle = \int_0^{\infty} d\omega \; \hat{C}_F(\omega) \int_0^{t} dt_1 \int_0^{t} dt_2 \; R(t-t_1) R(t-t_2) \cos \left[ \omega \left(t_1-t_2 \right)
\right].
\end{equation}
\end{widetext}
We introduce new integration variables $\tau=t-t_1$ and $u= t-t_2$ and  convert equation (\ref{pTpT}) into the form 
\begin{widetext}
\begin{equation} \label{p2}
\langle p^2(t)\rangle = \int_0^{\infty} d\omega \; \hat{C}_F(\omega) \int_0^{t} d\tau \int_0^{t} du \; R(\tau) R(u) \cos \left[ \omega \left(\tau-u \right) \right]. 
\end{equation}
\end{widetext}
We perform the limit  $t\to\infty$ to derive an expression for the average kinetic energy in the equilibrium state, namely,   
\begin{equation} \label{Elim}
E_k = \lim_{t \to \infty} \frac{1}{2M} \langle p^2(t)\rangle =
\frac{1}{2M} \int_0^{\infty} d\omega \; \hat{C}_F(\omega) I(\omega), 
\end{equation}
where 
\begin{align} \label{I}
I(\omega) &= \int_0^{\infty} d\tau \int_0^{\infty} du \; R(\tau) R(u) \cos \left[ \omega \left(\tau-u \right) \right] \nonumber \\ &= \hat{R}_L(i\omega) \hat{R}_L(-i\omega)
\end{align}
is the product of a Laplace transform of the response function $R(t)$. At this point, we can exploit the fluctuation-dissipation relation (\ref{d-f}) (Appendix C) to express the noise correlation spectrum $\hat{C}_F(\omega)$ by the dissipation spectrum $\hat{\gamma}_F(\omega)$ and convert (\ref{Elim}) to the form 
\begin{equation}\label{Ek7}
E_k =  \int_0^{\infty} d\omega \; \frac{\hbar \omega}{4M} \coth\left({\frac{\hbar \omega}{ 2k_BT}}\right) \;\hat{\gamma}_F(\omega)\hat{R}_L(i\omega) \hat{R}_L(-i\omega). 
\end{equation}
We observe that 
\begin{equation}\label{ho2}
\mathcal{E}_k(\omega) = \frac{\hbar \omega}{4} \coth\left({\frac{\hbar \omega}{ 2k_BT}}\right) 
\end{equation} 
is averaged (thermal) kinetic energy per one degree of freedom of the thermostat consisting of free harmonic oscillators \cite{feynman}. The remaining part of the integrand in Eq. (\ref{Ek7}) reads 
%
\begin{eqnarray}\label{P2}
\mathbb{P}(\omega) &=& \frac{1}{M}  \hat{\gamma}_F(\omega)  
\hat{R}_L(i\omega) \hat{R}_L(-i\omega) \nonumber \\ &=&
\frac{M}{\pi} \frac{\hat{\gamma}_L(i\omega) + \hat{\gamma}_L(-i\omega)}{[\hat{\gamma}_L(i\omega) +iM\omega] [  \hat{\gamma}_L(-i\omega) -iM\omega]}   \nonumber\\
&=& \frac{1}{\pi} \left[\hat{R}_L(i\omega) + \hat{R}_L(-i\omega) \right], 
\end{eqnarray}
%
where we used Eq. (\ref{RL2}) for $\hat{R}_L(z)$ and the relation between the Laplace and cosine Fourier transforms.  With these two expressions for $\mathcal{E}_k(\omega)$ and $\mathbb{P}(\omega)$, the final form of  the averaged kinetic energy $E_k$  of the Brownian particle reads 
\begin{equation}\label{Ek2}
E_k =  \int_0^{\infty} d\omega \; \mathcal{E}_k(\omega)\mathbb{P}(\omega).   
\end{equation}
\section{Fluctuation-dissipation relation}
We assume the factorized initial state  of the composite  system, i.e., $\rho(0)=\rho_S\otimes\rho_E$, where $\rho_S$ is an arbitrary state of the Brownian particle and $\rho_E$ is an equilibrium canonical state of the thermostat 
of temperature $T$, namely,
\begin{eqnarray}
\rho_E = \mbox{exp}(-H_E/k_B T)/\mbox{Tr}[\mbox{exp}(-H_E/k_B T)],
\end{eqnarray}
where:
\begin{equation}
H_E= \sum_i \left[ \frac{p_i^2}{2m_i} + \frac{1}{2} m_i \omega_i^2 q_i^2 \right] 
\end{equation}
is the Hamiltonian of the thermostat. The factorization means that there are no initial correlations between the particle and  the thermostat. The initial preparation turns the force $\eta(t)$ into the operator-valued quantum thermal  noise which in fact is a family of non-commuting operators whose commutators are $c$-numbers. This noise is unbiased and its mean value is zero, 
\begin{eqnarray}
\langle \eta(t) \rangle \equiv \mbox{Tr} \left[ \eta(t) \rho_E\right] = 0.  
\end{eqnarray} 
Its symmetrized correlation function 
\begin{equation} \label{correl}
C(t, u)=  \symkowm{\eta(t)}{\eta(u)}  = \frac{1}{2} \langle   \eta(t)\eta(u)+\eta(u)\eta(t)\rangle
\end{equation}
depends on the time difference 
\begin{align}\label{corr-f}
C(t, u) &= C(t - u) \nonumber \\ &= \sum_i \frac{\hbar c_i^2}{2 m_i \omega_i} \coth \left(\frac{\hbar \omega_i}{2k_B T} \right) \cos[\omega_i(t-u)] \nonumber\\
&=
 \int_0^{\infty} d \omega \, \frac{\hbar \omega}{2} \coth \left(\frac{\hbar \omega}{2k_B T}\right) J(\omega)\cos[\omega(t-u)], 
\end{align} 
where the spectral function $J(\omega)$ is given by Eq. (\ref{spectral}). 
The higher order correlation functions are expressed  by $C(t_i-t_j)$  and have the same form as statistical characteristics for classical stationary Gaussian stochastic processes. Therefore $\eta(t)$  defines  a quantum stationary Gaussian process with time homogeneous correlations. 

The dissipation and correlation functions  can be presented as cosine  Fourier transforms 
\begin{subequations}
\begin{align} \label{gatau}
&\gamma(t) = \int_0^{\infty} d \omega \,\hat{\gamma}_F(\omega) \cos(\omega t), \\ 
&C(t) = \int_0^{\infty} d \omega \,\hat{C}_F(\omega)\cos(\omega t), \label{C}
\end{align}
\end{subequations}
with their inverse 
\begin{subequations}
\begin{align} \label{inverseG}
\hat\gamma_F(\omega) = \frac{2}{\pi} \int_0^{\infty} dt \,\gamma(t) \cos(\omega t), \\
\hat{C}_F(\omega) = \frac{2}{\pi} \int_0^{\infty} dt \,C(t)\cos(\omega t), \label{inverseC}
\end{align}
\end{subequations}
If we compare Eqs. (\ref{diss}) and (\ref{corr-f})-(\ref{C}) then we observe that   
\begin{equation} \label{d-f}
\hat{C}_F(\omega)= \frac{\hbar \omega}{2} \coth \left(\frac{\hbar \omega}{2k_B T} \right) \,\hat{\gamma}_F(\omega).
\end{equation}
This relation between the spectrum $\hat{\gamma}_F(\omega)$ of dissipation and the  spectrum $\hat{C}_F(\omega)$ of thermal noise correlations is the body of the fluctuation-dissipation theorem \cite{call,kub} in which  quantum effects are incorporated via the prefactor in r.h.s. of Eq. (\ref{d-f}). We want to pay attention that the definition (\ref{spectral}) of the spectral density $J(\omega)$ differs from another frequently used form $\tilde J(\omega) = \omega J(\omega)$.  We prefer the definition (\ref{spectral}) because of a direct relation to the Fourier transforms of (\ref{diss}) and (\ref{gatau}), i.e. $J(\omega) = \hat{\gamma}_F(\omega)$. Here, the Ohmic case corresponds to $J(\omega)=$const.

For a finite number of the thermostat oscillators, all dynamical quantities are almost periodic functions of time, in particular the dissipation function $\gamma(t)$ and the correlation function $C(t)$. In the thermodynamic limit, when a number of oscillators tends to infinity, the dissipation function $\gamma(t)$ decays to zero as $t\to\infty$ and the singular spectral function $J(\omega)$  defined by  Eq. (\ref{spectral}) tends to a (piecewise) continuous function.   In such a point of view, dissipation mechanism is determined by the memory kernel  $\gamma(t)$ or equivalently by the spectral density of thermostat modes $J(\omega)$ which contains necessary  information on the particle-thermostat interaction.


\begin{thebibliography}{99}

\bibitem{huang} K. Huang, \textit{Statistical mechanics} (Wiley, New York, 1987)

\bibitem{ter} Y. P. Terletski\'i \textit{Statistical Physics} (North-Holland, Amsterdam, The Netherlands, 1971) 

\bibitem{hakim} V. Hakim, V. Ambegaokar, \textit{Phys. Rev. A} 32, 423 (1985)

\bibitem{ford85} G. W. Ford, J. T. Lewis, R. F. O'Connell, \textit{Phys. Rev. Lett.} 55, 2273 (1985)

\bibitem{ford88} G. W. Ford, J. T. Lewis, R. F. O'Connell, \textit{Ann. Phys. (N.Y.)} 185, 270 (1988)

\bibitem{ingol1} H. Grabert, P. Schramm, G. L. Ingold, \textit{Phys. Rep.} 168, 115 (1988)

\bibitem{weis} U. Weiss, \textit{Quantum Dissipative Systems} (World Scientific: Singapore, 2008)

\bibitem{landau} L. D. Landau, E. M. Lifshitz \textit{Statistical Physics, Part 1} (Butterworth-Heinemann, 3rd ed., 1980)

\bibitem{zubarev} D. N. Zubarev, \textit{Nonequilibrium statistical thermodynamics} (New York, Consultants Bureau, 1974)

\bibitem{breuer} H. P. Breuer and F. Petruccione, \textit{The theory of open quantum systems} (New York, Oxford University Press, 2002).  

\bibitem{grela} J. Grela, S. N. Majumdar and G. Scherer, \textit{Phys. Rev. Lett.} 119, 130601 (2017)

\bibitem{lewenstein} P. Massignan, A. Lampo, J. Wehr and M. Lewenstein, \textit{Phys. Rev. A} 91, 033627 (2015)

\bibitem{korbicz} J. Tuziemski and J. K. Korbicz, \textit{EPL} 112, 40008 (2015)

\bibitem{smirne} L. Ferialdi and A. Smirne, \textit{Phys. Rev. A} 96, 012109 (2017)

\bibitem{editor} D. Boyanovsky and D. Jasnow, \textit{Phys. Rev. A} 96, 062108 (2017)

\bibitem{ankerhold} B. Jack, J. Senkpiel, M. Etzkorn, J. Ankerhold, Ch. Ast and K. Kern, \textit{Phys. Rev. Lett.} 119, 147702 (2017)

\bibitem{carlesso} M. Carlesso, A. Bassi, \textit{Phys. Rev. A} 95, 052119 (2017)

\bibitem{lampo} S. H. Lim, J. Wehr, A. Lampo, M. A. Garica-March, and M. Lewenstein, \textit{J. Stat. Phys.} 170, 351 (2018)

\bibitem{china} H. Z. Shen, S. L. Su, Y. H. Zhou and X. X. Yi, \textit{Phys. Rev. A} 97, 042121 (2018)

\bibitem{lampo2} A. Lampo,  C. Charalambous, M. A. Garc{\'i}a-March, and M. Lewenstein, \textit{Quantum} 1, 30 (2018)  

\bibitem{bialas} P. Bialas and J. {\L}uczka, \textit{Entropy} 20, 123 (2018)

\bibitem{arxiv2018} P. Bialas, J. Spiechowicz and J. {\L}uczka, arXiv Preprint at arXiv:1805.04012 (2018)

\bibitem{maga} V. B. Magalinskij, \textit{J. Exptl. Theoret. Phys.} 36, 1942 (1959) 
[Sov. Phys. JETP 9, 1381 (1959)].

\bibitem{uler} P. Ullersma, \textit{Physica} 32, 27 (1966)

\bibitem{caldeira} A. O. Caldeira and A. J. Leggett, \textit{Ann. Phys. (N.Y.)} 149, 374 (1983); \textit{Ann. Phys. (N.Y.)} 153, 445 (1984)

\bibitem{ford} G. W. Ford, M. Kac, \textit{J. Stat. Phys.} 46, 803 (1987)

\bibitem{gaussian} P. De Smedt, D. D\"urr, J. L. Lebowitz, \textit{Commun. Math. Phys.} 120, 195 (1988)

\bibitem{van} N. Van Kampen, \textit{J. Mol. Liq.} 71, 97 (1977) 

\bibitem{et2} G. W. Ford, J. T. Lewis, R. F. O'Connell,  \textit{Phys. Rev. A} 37, 4419 (1988)

\bibitem{ph} P. H\"anggi, G. L. Ingold, \textit{Chaos} 15, 026105 (2005)

\bibitem{chaos} J. {\L}uczka, \textit{Chaos} 15, 026107 (2005)


\bibitem{feynman} R. P. Feynman, \textit{Statistical Mechanics} (Westview Press, USA, PA, 1972) 

\bibitem{anomal1} R. Morgado, F. A. Oliveira, G. G. Batrouni and 
A. Hansen, \textit{Phys. Rev. Lett.} 89, 100601 (2002) and refs therein

\bibitem{anomal2} S. A. McKinley, H. D. Nguyen, \textit{SIAM J. Math. Anal.} 50, 5119 (2018) 

\bibitem{zwan} R. Zwanzig, \textit{J. Stat. Phys.} 9, 215 (1973)

\bibitem{call} H. B. Callen and T. A. Welton, \textit{Phys. Rev.} 83, 34 (1951)

\bibitem{kub} R. Kubo, \textit{Rep. Prog. Phys.} 29, 255 (1966)



\end{thebibliography}
\end{document}